\documentclass[12pt]{iopart}
\usepackage{iopams}  
\usepackage{graphicx}
\def\<{\langle} \def\>{\rangle} \def\(({\left(} \def\)){\right)}
\def\[[{\left[} \def\]]{\right]}

\newcommand{\be}{\begin{equation}} \newcommand{\ee}{\end{equation}}
\newcommand{\bea}{\begin{eqnarray}} \newcommand{\eea}{\end{eqnarray}}

\begin{document}

\title[The nature of spin-glass phases in 2D discrete spin glasses]{The nature of the different zero-temperature phases in discrete
  two-dimensional spin glasses: Entropy, universality, chaos and
  cascades in the renormalization group flow}

\author{Thomas J\"org and Florent Krzakala}
\address{Laboratoire
  PCT, UMR Gulliver CNRS-ESPCI 7083, 10 rue Vauquelin, 75231 Paris,
  France } 

\begin{abstract}The properties of discrete two-dimensional spin glasses
  depend strongly on the way the zero-temperature limit is taken.  We
  discuss this phenomenon in the context of the Migdal-Kadanoff
  renormalization group. We see, in particular, how these properties
  are connected with the presence of a cascade of fixed points in the
  renormalization group flow. Of particular interest are two unstable
  fixed points that correspond to two different spin-glass phases at
  zero temperature.  We discuss how these phenomena are related with
  the presence of entropy fluctuations and temperature chaos, and
  universality in this model.
\end{abstract}


\newpage

Since the beginning of the study of disordered systems with
renormalization and scaling methods the question of universality and
of the relevance of the realization of the disorder have been a key
issue \cite{imry:75}. Why indeed should we accept that some abstract
models are the archetype of a very broad class of systems if their
behavior depends drastically on tiny details?  The Edwards-Anderson
\cite{edwards:75} model is one of the models that are widely regarded
as a prototype of disordered systems in statistical physics. It is an
Ising model with disordered and competitive interactions and has been
the source of many surprises and developments in the last thirty years
\cite{mezard:87,mezard:10}.

Its two-dimensional (2D) version, one of its simplest settings, has
very special properties with respect to universality. It is now agreed
that the spin-glass phase exists only at zero temperature
\cite{bhatt:88,saul:93,houdayer:01}, where the spin-glass susceptibility
diverges. However, the behavior of the model seems to depend
drastically on microscopic details and in particular discrete and
continuous couplings leads to different properties
\cite{hartmann:01b,amoruso:03} at zero temperature.  On the other
hand, for nonzero temperatures, strong evidence for universal critical
behavior has been observed
\cite{joerg:06a,lukic:06,toldinPelissetto:10}. This rises questions on
the very nature of universality, if any, in strongly disordered
bidimensional systems. Why such a difference? What are the mechanisms
behind this behavior? The key to understand these features lies in the
difference between strictly zero and vanishing temperature
\cite{krzakala:00b,joerg:08b}. When temperature is finite, entropy
fluctuations play a major role and, for large enough sizes, bring back
discrete models to the continuous class \cite{joerg:06a}. These
mechanisms were further exploited in \cite{huse}, with a special
emphasis on the so-called temperature chaos effect \cite{fisher:86,bray:86,bray:87}.

In this Letter we study the behavior of continuous and discrete 2D
spin glasses in the context of the Migdal-Kadanoff renormalization
group (MKRG) \cite{Kadanoff}. We observe that indeed continuous and
discrete $2D$ spin glasses are eventually associated with the very
same physically relevant fixed point. However, there are also key
differences that are the consequence of a cascade of two repulsive
fixed points in the MKRG flow in the case of the discrete model. In a
real-space picture, these phenomena are associated with two different
temperature-dependent crossover length scales, so that two different
zero-temperature spin-glass phases can be observed depending on the
system size $L$ and the temperature $T$.

This Letter is organized as follows: first, we define the model and
recall briefly the MKRG. We then study the MKRG flow and observe one
attractive paramagnetic fixed point and two repulsive ones
corresponding to the $T=0$ discrete and continuous classes,
respectively. We discuss the crossover length scales between different
regimes, the cascade of fixed points and its effects on effective
critical exponents which may easily camouflage the fact that the
critical behavior of the 2D spin glass is universal.

\section{Two-dimensional spin glasses}
The Hamiltonian of the Edwards-Anderson Ising spin
glass\cite{edwards:75} is given by
\begin{equation} {\mathcal H} = - \sum_{<i, j>} J_{ij} S_i S_j.
  \label{eq:ham}
\end{equation}
The spins $S_i$ lie on a square lattice of size $N = L^2$ in two space
dimensions and the interactions between the spins are between nearest
neighbors. We shall consider two different models: in the {\it
  discrete} one\footnote{In this letter, discrete means discrete
  energy spectrum, thus not allowing for irrational discrete values
  for the coupling $J_{ij}$ as in \cite{amoruso:03}.}, the
interactions are chosen randomly as $J_{ij} \in \{\pm 1\}$ while in
the {\it continuous}, or {\it Gaussian} model, we shall use a Gaussian
distribution of the $J_{ij}$ with zero mean and unit variance,
instead.

It is by now generally accepted (although there is no rigorous proof
of this, but see\cite{nishimori:09}) that there is no spin-glass
transition at finite temperature in both cases. Instead, the
spin-glass susceptibility diverges when $T \to 0$ and the spin-glass
phase exists only at zero temperature.  At any {\it finite}
temperature, however, this implies the existence of an equilibrium
length scale $\xi_{\rm eq}(T)$ beyond which the spin-glass correlation
decays exponentially fast and the system is effectively in a
paramagnetic state. For length scales below $\xi_{\rm eq}(T)$, the
correlation function decays only as a power-law and the system has a
spin-glass-like correlation. If one is looking at a system of size $L
\ll \xi_{\rm eq}(T)$, it thus looks very much as a spin glass (it has
power-law correlations) and only for larger sizes, when $L \gg
\xi_{\rm eq}(T)$ does the system finally look paramagnetic.

\section{Migdal-Kadanoff renormalization group}
Low-dimensional systems are often well described by the
Migdal-Kadanoff renormalization group and 2D spin glasses are no
exception \cite{Kadanoff,southern:77,fisher:91c,migliorini:98}. We shall thus work within the MKRG
approach and study how different are 2D discrete and continuous spin
glasses. For a disordered system such as a spin glass, the MKRG is a
{\it functional} renormalization group method, as the quantity that is
being renormalized is the distribution of couplings $P(\beta
J)$. Starting from a given inverse temperature $\beta=1/T$ and an
initial distribution of couplings $P_{\rm init}(\beta J)$, MKRG allows
to follow the (approximate) flow under renormalization (for a detailed
description see for instance \cite{southern:77,fisher:91c,migliorini:98}). There are many ways to write
the MKRG recursion for 2D spin glasses. We shall follow
\cite{amoruso:03} and use the recursion on a hierarchical lattice with
$b=3$ branches and $s=3$ spins per branch yielding a model with
effective dimension $D=2$:
\be P^{G+1}(\beta J)\!\!=\!\!\int\!\! \prod_{i=1}^9 \((d \beta_i
P^G(\beta J_i)\)) \delta{[\beta J\!-\!{\cal{F}} (\{{\beta
    J_i\}}_{i=1\ldots9})]}
\label{MKRG}
\ee \bea \mbox{with~~~}{\cal{F}} (\{{\beta
  J_i\}}_{i=1\ldots9})&=&\mbox{atanh}({\tanh{\beta J_1} \tanh{\beta
    J_2}\tanh{\beta J_3} } )\phantom{.}
\nonumber \\
&+&\mbox{atanh}({\tanh{\beta J_4} \tanh{\beta J_5}\tanh{\beta J_6}
})\phantom{.}
\nonumber \\
&+& \mbox{atanh}({\tanh{\beta J_7} \tanh{\beta J_8}\tanh{\beta J_9} }).
\nonumber \eea
This recursion can be implemented trivially using population
dynamics. It is also sometimes convenient to see the MKRG as an exact
solution on a hierarchical lattice \cite{mckay:82}. In this case one
interprets the ``time'' in the RG flow (that is, the index $G$ in
Eq.~\ref{MKRG}) as related to the effective size of the system via
$L=3^G$.  This allows to characterize length scales in the system, and
it also provides an exact realization of the droplet/scaling theory
\cite{fisher:86,bray:86,bray:87}, which is, at least for the $2D$ model, accepted to
be a good description.

\begin{figure}
  \hspace{-0.5cm}
  \includegraphics[width=1.1\columnwidth]{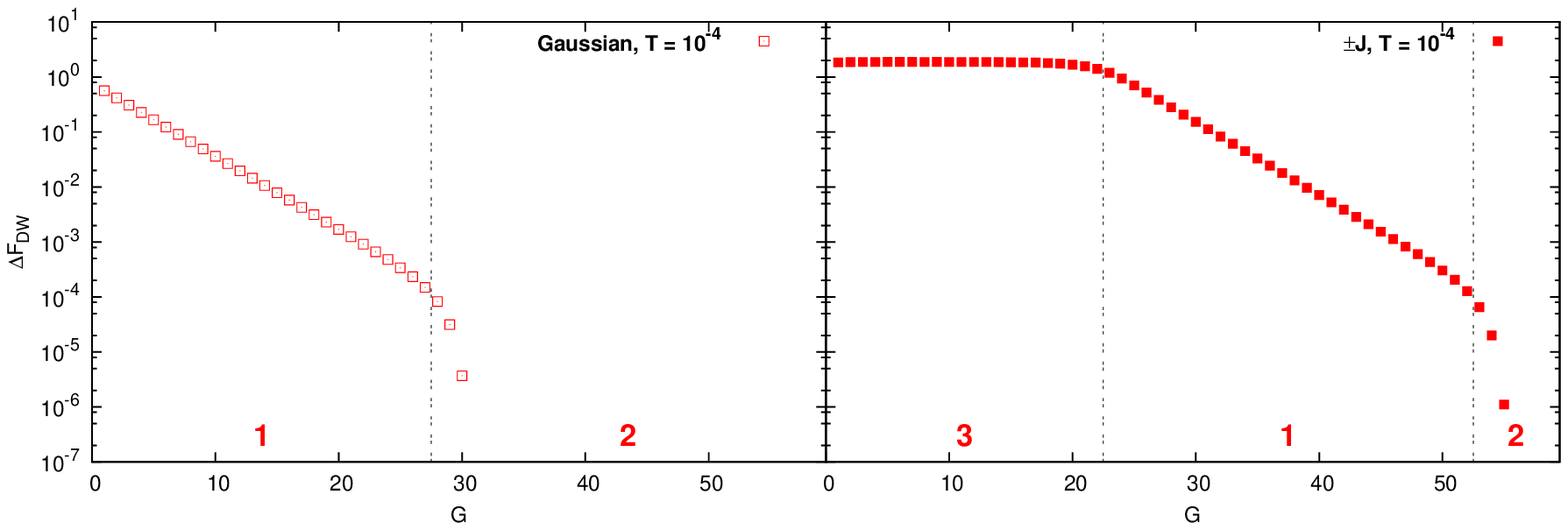}
  \center \includegraphics[width=0.5\columnwidth]{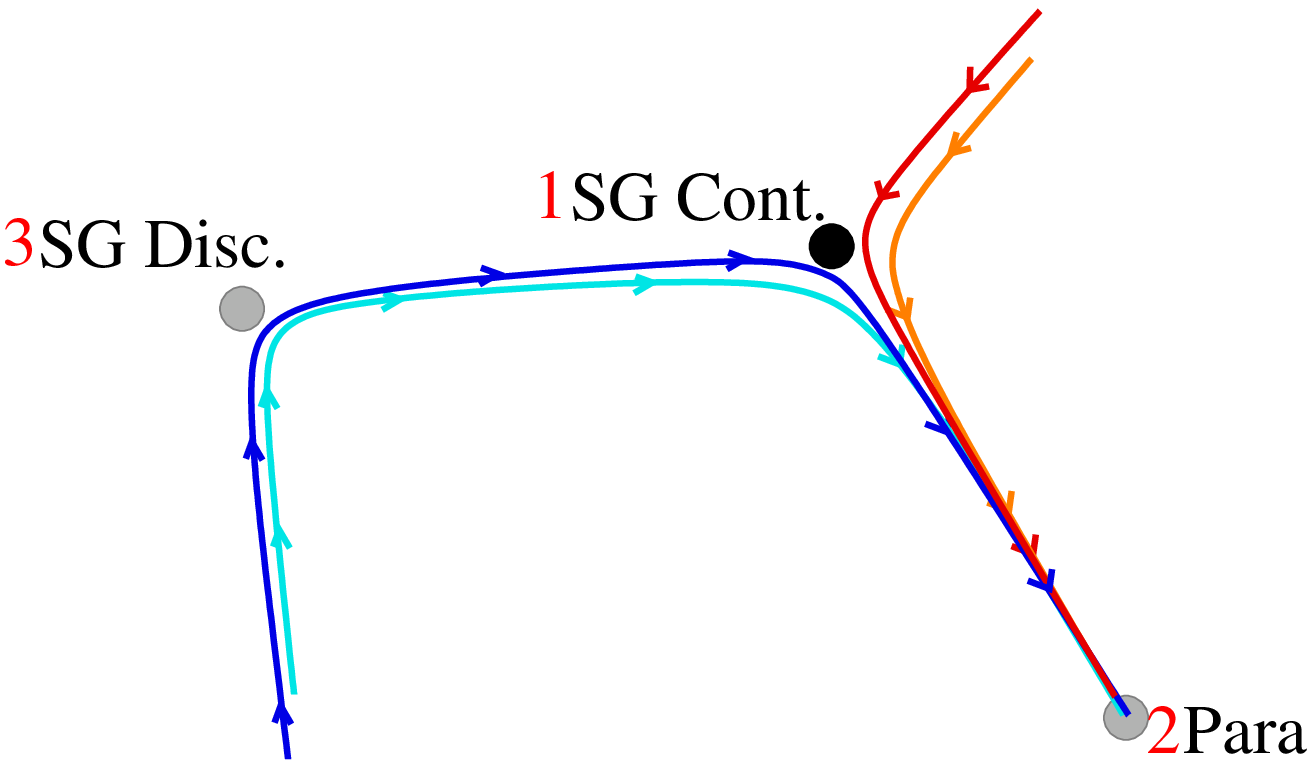}
 \caption{Top: The domain-wall free energy $\Delta F_{\rm DW}= \sqrt{\langle (J_{\rm p} - J_{\rm a})^2 \rangle}$ associated    with an effective bond after $G$ MKRG iterations (where the effective 
   size of the system is $L=3^G$), in the low-temperature phase of the 
   Gaussian (left) and binary (right) spin glasses. In the Gaussian 
   case one is in the spin-glass phase where $\Delta F_{\rm DW} \propto 3^{\theta G}$, 
   with $\theta \approx -0.278$ until the effective size of the system is larger
   than a crossover length $\xi_{\rm eq}(T)=T^{-\nu}$ where the system
   becomes paramagnetic and $\theta=-\infty$. In the discrete case, the
   system is first in a phase where $\theta^\prime=0$, until a first length scale
   $\ell_{{\rm c}}$ is crossed, and only at this point it decays with
   the continuous exponent $\theta \approx -0.278$. When the effective
   system is even larger, such that $L\gg L_{\rm G} \propto
   T^{-\nu_{\rm eff}}$, the system becomes again paramagnetic: there are
   therefore two different spin-glass phases in the zero-temperature
   limit. Bottom: the corresponding flow in the MKRG with the three
   different fixed points. Binary spin glasses are cascading through the
   three fixed points.
   \label{fig.1}}
\end{figure}

\section{Gaussian couplings}
When starting from a Gaussian distribution $P_{\rm {init}}(\beta J)$
at low temperature, the MKRG flows to a symmetric slightly
non-Gaussian distribution with zero mean and a variance decaying with
$G$ such that $\langle J^2 \rangle \propto L^{\theta}$, where
${\theta} \approx -0.278$. At strictly $T=0$, this is the
strong-coupling fixed point describing the continuous spin-glass
phase, but at any finite temperature $T$ the fact that $\theta<0$
implies that the bonds are not robust and that the spin-glass phase
does not survive (and that the fixed point is unstable). This is
precisely what happens once enough recursions are performed and the
MKRG flows to a trivial delta function exponentially fast: this is the
stable paramagnetic pseudo-fixed point where formally ${\theta} =
-\infty$ (see Fig.~\ref{fig.1} and Fig.~\ref{fig.2}.)

Following the standard droplet interpretation, scaling relations can
be obtained by realizing that the thermodynamic behavior is dominated
by large excitations (the droplets) of size $\ell$ and $O(1)$-energy
that can be found with a probability scaling as $\ell^{\theta}$. The
probability that two spins at distance $r$ are correlated thus decays
as $r^{\theta}$ and this will be on the order of the thermal
fluctuations when $r$ reaches the equilibrium length scale $\xi_{\rm
  eq} \propto T^{1/\theta}$. In other words the correlation function
behaves as $C_{\rm Gauss}(r,T) \propto r^{\theta} e^{-r/\xi_{\rm
    eq}(T)}$ which indicates that the critical exponent for the
correlation length is simply $\nu=-1/\theta$. Indeed, when performing
the MKRG at finite $T$, we observe that when the effective size $L$ is
of the order of $\xi$ (which eventually happens for any positive
temperature, provided we do enough recursions), the value of $\theta$
drops from $-0.278$ to $-\infty$.  This is the sign that we are now in
the paramagnetic phase and that there are no correlations beyond the
scale $\xi_{\rm eq}(T)$ (see again Fig.~\ref{fig.1}).

We recall all the MKRG exponents for the Gaussian model, together with
the estimated ones from the $2D$ square lattice in Table
\ref{Table_GAUSS}.  The values are very close, showing that the
critical properties are well captured by the MKRG approximation.

\begin{table}
  \begin{center}
    \begin{tabular}{l|c|c|}
      & MK     &  2D \\
      \hline
      $\theta$ & -0.278 & {-0.287} \cite{hartmann:02c} \\
      \hline
      $\nu=-1/\theta$ & 3.597 & 3.48  \\
      \hline
      $d_s$ & 1 & 1.28 \cite{amoruso:06a,Bernard:07}\\
      \hline
      $\zeta=\frac {d_S}2 - \theta $& 0.778 & 0.927 \\
      \hline
    \end{tabular}
    \caption{Gaussian couplings: critical exponents from the MKRG
      approximation and the 2D square lattice. \label{Table_GAUSS}}
  \end{center}

\end{table}

\section{Discrete spin glasses}
The discrete $\pm J$ model displays a more puzzling phenomenology.  It
was first realized that strictly at zero temperature, one finds
$\theta^\prime=0$ (we shall use a prime for all exponents in the
discrete model that are related with this fixed point, that we will
refer to as the {\it discrete fixed point}), both in MKRG
\cite{amoruso:03} and in the 2D model \cite{hartmann:01b}, suggesting
a different universality class with respect to Gaussian
disorder. However, it was soon realized that a small perturbation
(such as discrete, but fractional couplings) were enough to
destabilize this fixed point, and the MKRG was then flowing to the
continuous one \cite{amoruso:03} where $\theta=- 0.287$. Subsequent
Monte-Carlo simulations of the 2D model indicated that some
observables were in the same universality class in both the discrete
and continuous models\cite{joerg:06a}. This rises questions: how one
goes from different results at zero temperature in the discrete and
the continuous model to the same ones at finite $T$?  We will see that
there are actually two different unstable fixed points, corresponding
to two different spin-glass phases, that can be observed depending on
how the $T=0$ limit is taken, and a stable pseudo-fixed point
corresponding to the paramagnetic phase.

\subsection{The zero-temperature discrete phase} 
We first repeat the $T=0$ analysis of \cite{amoruso:03} and, indeed,
find $\theta^\prime=0$. That means that the $P^G(J)$ (which we rewrite
in order to remove the dependence in $\beta$ by taking formally the
limit $\beta \to \infty$ in Eq.~\ref{MKRG}) converges to a nontrivial
function that does not evolve under renormalization.  This fixed point
thus describes a zero-temperature spin-glass phase, whose properties
are different from the ones seen in the Gaussian model.

The nontrivial phenomenon, however, is that the flow is {\it not}
going directly to the paramagnetic pseudo-fixed point, but instead it
is {\it first} governed by the continuous fixed point, and {\it then}
finally flows to the paramagnetic one (see Fig.~\ref{fig.1}) implying
a nontrivial cascade in the MKRG flow.

\begin{table}
  \begin{center}
    \begin{tabular}{l|c|c|}
      & MK & 2D\\
      \hline
      $\theta^\prime$ & 0 & 0 \\
      \hline
      $\nu^\prime = 1/\zeta^\prime$ & 2.597 & 1.43 \ldots 1.83 \\
      \hline
      $d_s^\prime$ & 0.77 & 1.095 \ldots 1.395 \cite{melchert:07} \\
      \hline
      $\zeta^\prime=\frac {d_S^\prime}2 $& 0.385 & 0.5 \ldots  0.7 \\
      \hline\hline
      $\nu_{\rm eff} = \nu + \nu^\prime$ & 6.195 & 4.9 \ldots 5.5 \\
      \hline
    \end{tabular}
    \caption{Discrete $\pm J$ couplings: critical exponents associated
      with the ``discrete'' fixed point in the MKRG approximation and
      the estimated ones from the 2D square lattice. The exponent $\nu_{\rm eff}$
      describes the divergence of the spin-glass correlation length using
      a scaling in the system size $L$ and the bare temperature $T$. 
      \label{Table_DISCRETE}}
  \end{center}
\end{table}

\begin{figure*}[htb]
  \hspace{-1.2cm}
  \includegraphics[width=1.1\columnwidth]{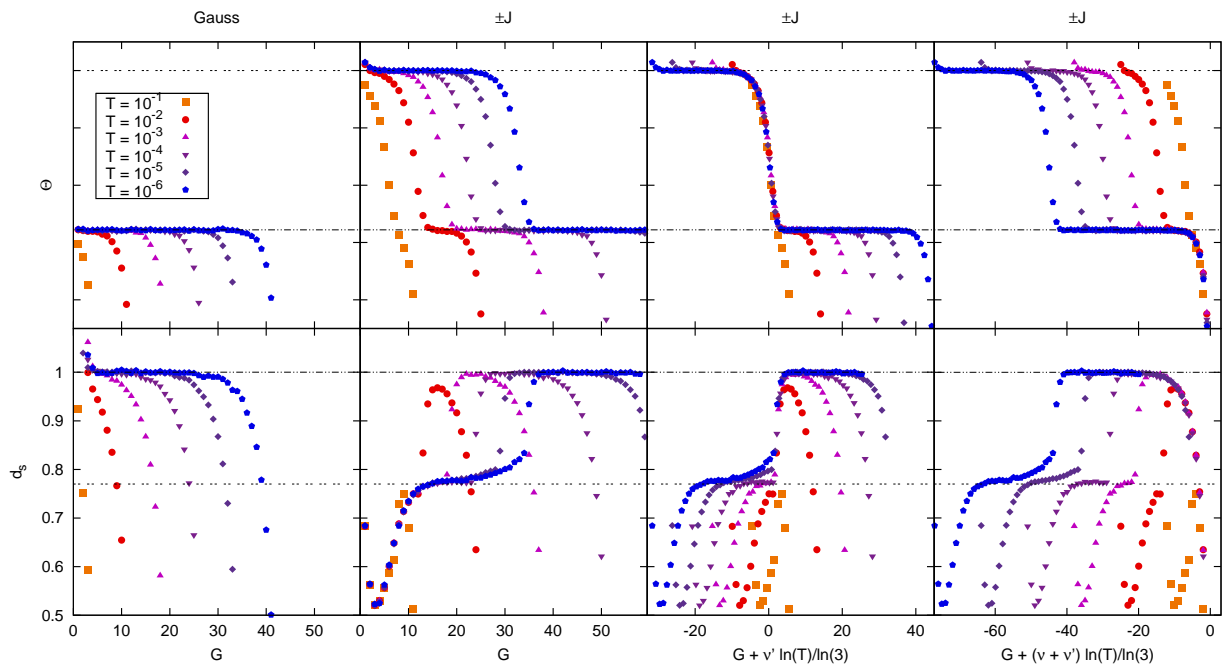}
  \caption{(Color online) Effective $\theta$ and fractal dimension of
    domain walls $d_s$ for the Gaussian and $\pm J$ models as a
    function of the number of generations $G$ in the MKRG for
    different temperatures $T$. The length scale probed by the MKRG is
    $L=3^G$. For the discrete model, the rescaling using the exponent
    $\nu^\prime = 1/\zeta^\prime$ allows to superimpose the curves in
    the crossover region from the discrete to the continuous fixed
    point, while using $\nu_{\rm eff}=\nu+\nu^\prime$ one can
    superimpose the crossover from the continuous fixed point to the
    paramagnetic region.  \label{fig.2}}
\end{figure*}

\begin{figure}[htb]
  \vspace{0.4cm}
  \includegraphics[width=1\columnwidth]{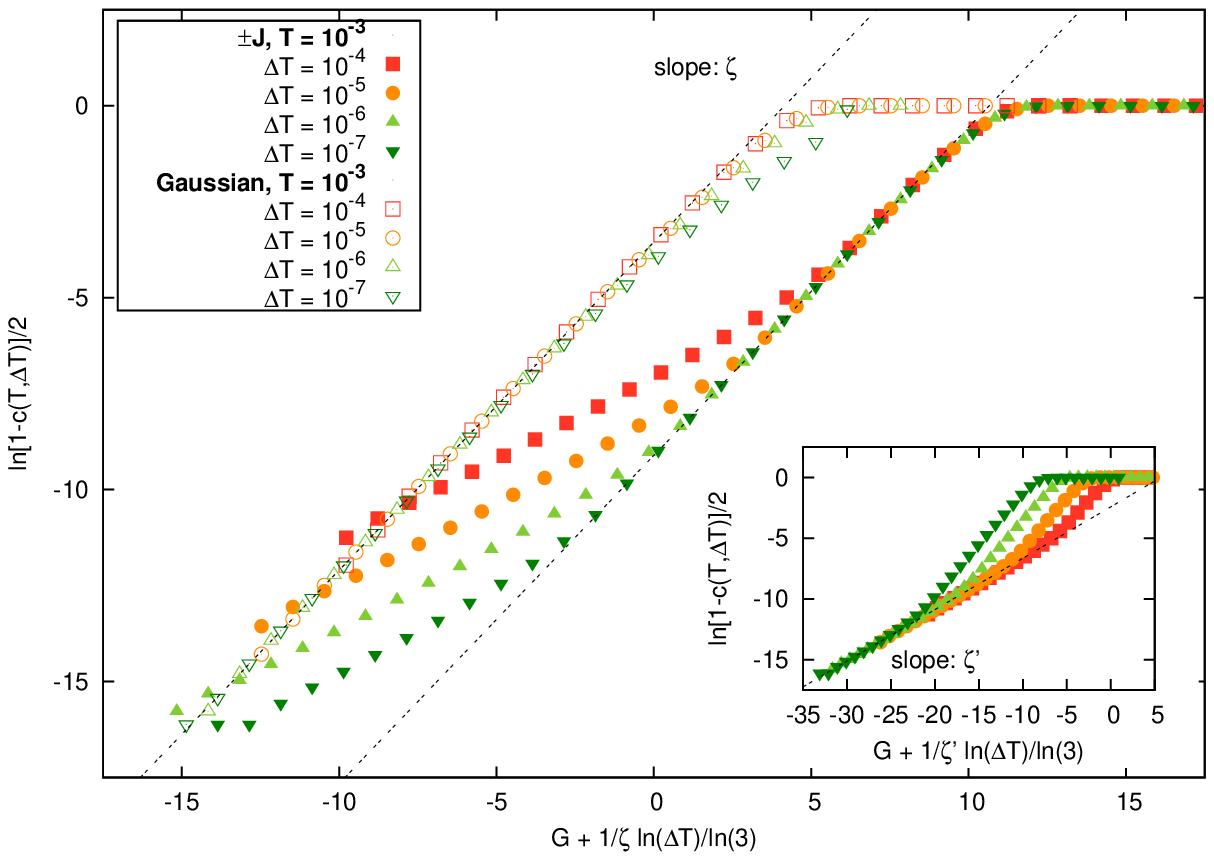}
  \caption{(Color online): Temperature chaos in 2D spin glasses: we
    show $\ln(1-C(T,\Delta T))/2$ where $C(T,\Delta T)$ is the
    correlation between the couplings at temperature $T$ and $T+\Delta
    T$. In the Gaussian model, the system is characterized by a
    critical exponent $\zeta$, while in the binary case one can
    observe either the exponent $\zeta$ or $\zeta^\prime$ depending on
    the values of $L$ and $T$. \label{chaos}}
\end{figure}

\subsection{Entropic fluctuations and temperature chaos}
The reasons behind the instability of the discrete fixed point are, as
first discussed by \cite{joerg:06a}, entropic fluctuations that at any
finite temperature make the zero-temperature and the
vanishing-temperature limits different (a very similar phenomenon
appears in diluted $3D$ spin glasses \cite{joerg:08b}, where the
critical dilution calculated by first setting the temperature to zero
and then taking the system size to infinity is strictly larger than
when first taking the system size to infinity at finite temperature
and then taking the temperature to zero). This is in fact deeply
connected with the notion of temperature
chaos\cite{huse,fisher:86,bray:86,bray:87,mckay:82,krzakala:02,nifle:92}.

Equilibrium states of spin glasses are sensitive even to very small
perturbations. Let us repeat here the classical thermodynamic argument
of \cite{fisher:86,bray:87}. Consider a spin glass in the scaling/droplet
approach: take two equilibrium states at temperature $T<T_{{\rm c}}$
differing by a very large droplet of characteristic size $\ell$. Then
the two states have free energies that differ by $\Delta F(T)=\Delta E
- T\Delta S \approx \Upsilon(T)\ell^{\theta}$, where $\Upsilon(T)$ is
the energy stiffness coefficient. When one changes the temperature by
$\Delta T$, then $\Delta F (T + \Delta T) \approx \Delta E-(T+\Delta
T)\Delta S$ so that $\Delta F(T + \Delta T) \approx \Upsilon(T)
\ell^{\theta} - \Delta T \Delta S$. In this phenomenological approach,
the entropy difference is associated with the droplet's surface so
that $\Delta S$ has a random sign and a typical magnitude
$\sigma(T)\ell^{d_{{\rm s}}/2}$, where $\sigma(T)$ is called the
entropy stiffness and $d_{{\rm s}}$ is the fractal dimension of the
droplet's surface. If $d_{{\rm s}}/2 > \theta$, which follows from
droplet theory, then $\Delta F(T + \Delta T)$ can change sign between
T and $T + \Delta T$ for length scales greater than
\begin{equation}
  \ell_{{\rm c}}\propto\bigg(\frac{\Upsilon(T)}
  {\sigma(T)\Delta T}\bigg)^{1/{\zeta}}
\end{equation}
with $\zeta=d_{{\rm s}}/2-\theta$. This can be checked in the MKRG
using the correlation between the effective couplings between the
border spins at different temperatures
\be C(T,\Delta T)=\frac{\langle J(T)J(T+\Delta T)\rangle}{\langle J(T)
  \rangle \langle J(T+\Delta T) \rangle}.  \ee

Besides the scaling in $L/\ell_{{\rm c}}$, it is also
expected\cite{nifle:92} that for small values of $\Delta T$ one has
$1-C \propto L^{2\zeta}$. As shown in Fig.~\ref{chaos} we find that
indeed one can rescale the curves and observe the predicted slope
using $\zeta=0.78$ for the Gaussian system. Using the definition of
$\zeta$, this shows that $d_s=1$ in MKRG, as is widely accepted. For
the discrete case, however, we first find a $\zeta^\prime \approx
0.385$ in the slope instead. This was to be expected since
$\theta^\prime \neq \theta$. In turns, this indicates that $d_s^\prime
\approx 0.77$. This is surprising, since in real systems $d_s$ cannot
be lower than $d-1$, and one has no choice but to accept it as a
strange byproduct of the MKRG approximation. The point, however, is
that once enough recursions are performed, the exponent in the slope
is not given by the discrete fixed point exponent $\zeta^\prime$
anymore, but by the continuous one $\zeta$, and the points superimpose
where rescaled by $\zeta$: this is a clear sign that we are now
effectively in the continuous spin-glass class (and this was in fact
first noticed by \cite{lukic:06}). We also find that $d_s^\prime$ is
smaller than $d_s$ in the MKRG as well as on the square lattice model.

In a real-space interpretation, one can picture the phenomenon by
thinking of droplets ``dressed'' with entropic fluctuations: when the
entropy plays a role, at finite temperature, the droplet's cost is
given by an interplay between the values of $J$, which are discrete,
but also by the entropic values proportional to $T$ (given by some
``spins fous'' in effective zero field \cite{krzakala:00b}) which is
continuous: the free-energy cost of an excitation starts to be
slightly continuous and the MKRG flow must leave the discrete fixed
point to go to the continuous fixed point. The fact that this is due
to entropy fluctuations, and therefore temperature chaos, is yet
another illustration of the its importance in spin glasses.

\subsection{The continuous spin-glass regime}
For sizes larger than $\ell_{\rm c}(T)$, we thus expect the MKRG flow
to be governed by continuous fixed point. Indeed, we recover the
``continuous class'' critical exponents $d_s$ and $\theta$ once the
number of iteration is large enough such that $L > \ell_{\rm c}$: this
can be seen directly in the behavior of the domain-wall exponent
(Fig.~\ref{fig.1} and Fig.~\ref{fig.2}) where the three different
regimes (discrete, continuous and paramagnetic) are observed
successively. We have also checked, by scaling the values of $\theta$
obtained after a given number of iterations $G$ at temperature $T$,
that the crossover from the discrete to the continuous fixed point
arises at $\ell_{\rm c}$, and is therefore controlled by the exponent
of the discrete fixed point $\nu^\prime = 1/\zeta^\prime$.

\subsection{The paramagnetic regime}
At any finite $T$, the flow must eventually end in the paramagnetic
attractive pseudo-fixed point. Thus, according to the MKRG, there
should be three different regimes: for sizes $L<\ell_{\rm c}$ we have
$\theta^\prime=0$. For $\ell_{\rm c}<L<L_{\rm eq}$ we have
$\theta=-0.278$ while for $L>L_{\rm eq}$ we are in the paramagnetic
phase and $\theta=-\infty$. The picture of the flow is one of a
cascade of three fixed points (two unstable and a final stable one,
see Fig.\ref{fig.1}). However, due to this cascade, the two repulsive
fixed points combine to produce an apparent nontrivial exponent for
the crossover finite-size length scale, which is given by $\nu_{\rm
  eff}=\nu+1/\zeta^\prime$, instead of $\nu$ in the Gaussian problem.
This can also be understood directly in the scaling picture: Let us
consider the probability that two spins at distance $r$ are coupled.
Starting from size $\ell_{\rm c}=T^{1/\zeta^\prime}$ it decays as
$p\propto (r/\ell_{\rm c})^{\theta}$. Indeed, we saw that the
domain-wall free-energy is $O(1)$ for $r=\ell_{\rm c}$. This indicates
that this will be of order $O(T)$ for $L \propto T^{1/\theta}
T^{-1/\zeta^\prime}$ so that $L_{\rm eq}\propto
T^{1/\theta-1/\zeta^\prime}$. In other words the effective exponent
using the system size and the bare temperature $T$ as scaling
variables is not $\nu$, but $\nu_{\rm eff}=\nu+1/\zeta^\prime$. We
will, however, see that the use of these scaling variables only
camouflages the fact that the divergence of the spin-glass correlation
length is simply given by the exponent $\nu$.

\section{Universality}
Universality is a tricky problem in this model, precisely because of
this nontrivial cascade of fixed points! Depending on the temperature
$T$ and the size $L$, one can be sensitive either to the
``continuous'' critical point, to the discrete one, or even a
combination of the two in simulations, and it is important not to
confuse the two. Once, however, the peculiar picture of the flow
(Fig.~\ref{fig.1}) is understood, it is clear that the critical fixed
point governing the RG flow before the paramagnetic is reached is the
same for both Gaussian and binary disorder.

\begin{figure}[htb]
  \includegraphics[width=0.485\columnwidth]{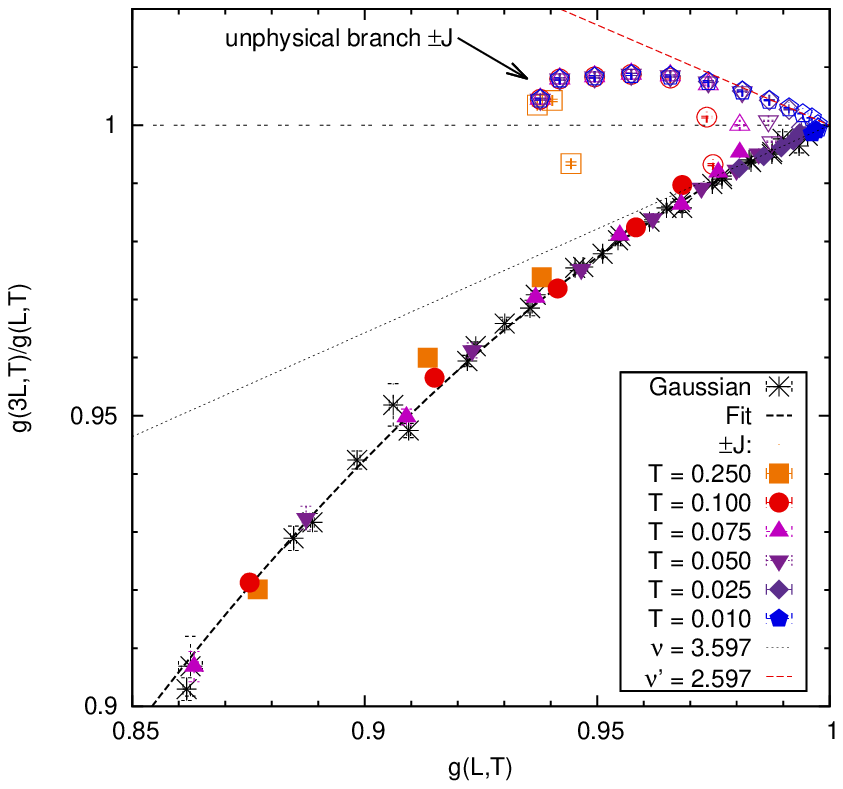}
  \includegraphics[width=0.455\columnwidth]{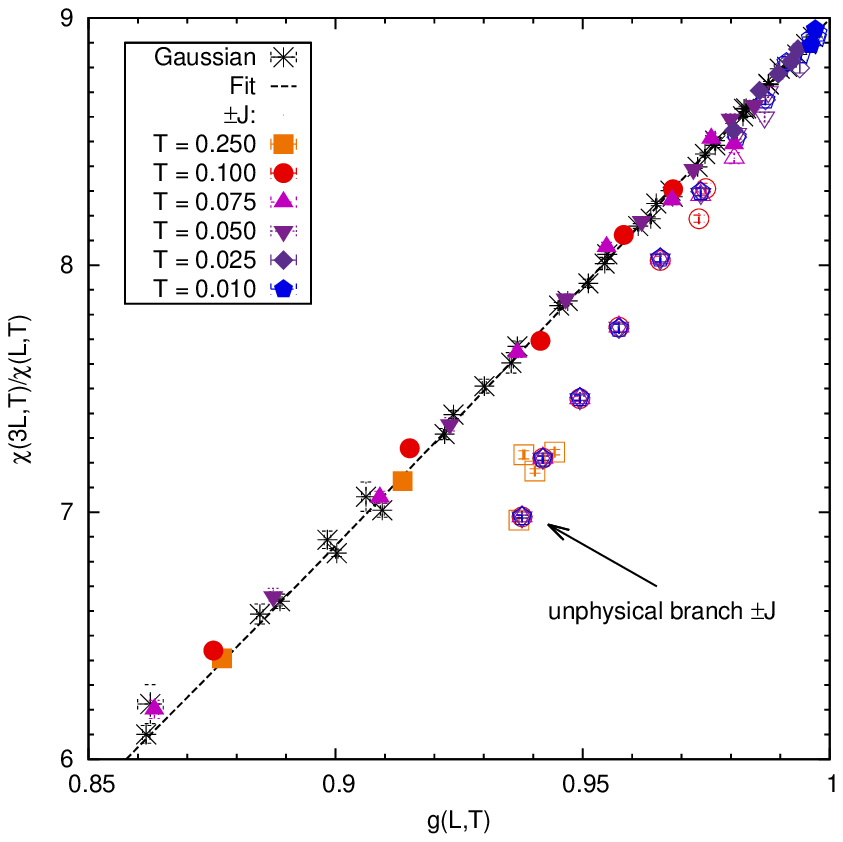}
  \caption{(Color online) In the left panel the scaling function of the
    Binder ratio $g(L,T)$ is shown for that Gaussian and $\pm J$
    model.  The data collapse shows that they share same scaling
    function and hence the same critical exponent $\nu$. In the right
    panel the scaling function of the spin-glass susceptibility $\chi$
    is shown. The data collapse on the physical branch of the scaling
    functions shows that the critical exponent $\eta$ is
    universal.\label{binder}}
\end{figure}

We shall now see how one can observe the continuous class in
simulation of the discrete model. As argued in
\cite{joerg:08c,toldinPelissetto:10} using bare (unrenormalized)
couplings (e.g., temperature) as scaling variables is not a good idea
in this model and leads to confusing results, but instead dressed
(renormalized) couplings (e.g., Binder ratio) should be used. This was
precisely the idea used in the simulations in $2D$ that observed
strong evidence for universality \cite{joerg:06a,toldinPelissetto:10}.
A good examples of such a quantity is the Binder cumulant
$g(L,T)=\frac{1}{2}\left(3 -\frac {\langle q^4 \rangle}{\langle q^2
    \rangle^2}\right)$ of the overlap distribution $P(q)$ and the
spin-glass susceptibility $\chi=N \langle q^2 \rangle$. These are
classical parameters in simulations and they can be computed in MKRG
as well (see \cite{moore:98} for details) which gives a good idea of
what is to be expected in the 2D system on a square lattice at finite
size. We display the scaling functions of the Binder ratio and the
spin-glass susceptibility as introduced in \cite{joerg:08c} for the
Gaussian and the discrete disorder in Fig.~\ref{binder}. These graphs
allow one -- amongst other things -- to determine the critical
exponents $\nu$ and $\eta$.  The points obtained from the Gaussian and
the discrete model superpose on a nontrivial universal curve, apart
from the fact that the discrete model has an additional branch which
is related to the discrete fixed point, but which, however, is 
irrelevant for the critical exponents observed in the thermodynamic 
limit. This is, indeed, precisely what is also observed in simulations
of the 2D binary square lattice model \cite{joerg:06a}. It is
interesting to note that, in contrast to the continuous fixed point,
the relation $\theta=-1/\nu$ does not hold for the discrete fixed
point where $\theta^\prime=0$, but $\nu^\prime=2.597$.

\begin{figure}[htb]
  \includegraphics[width=1\columnwidth]{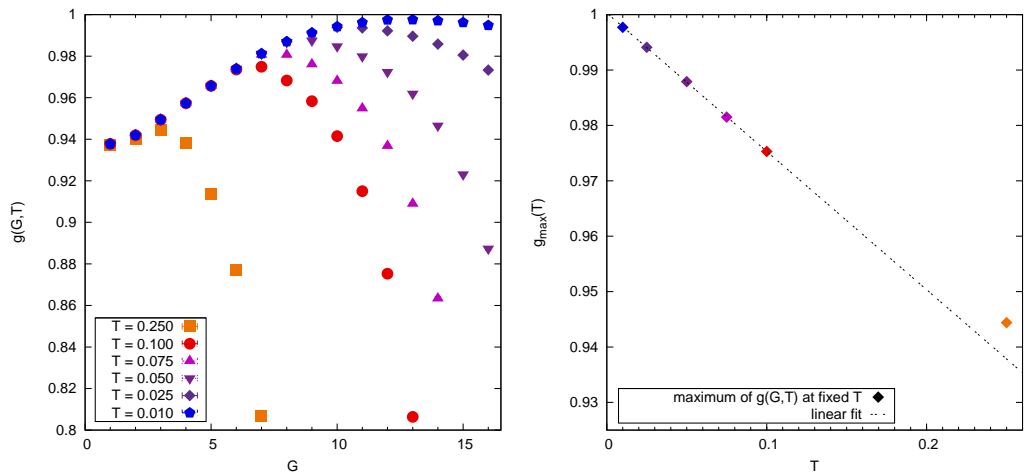}
  \caption{(Color online) In the left panel the Binder ratio $g(G,T)$ is shown as
    a function of the number of generations $G$ used in the construction of the 
    hierarchical lattice. In the right panel the maximal value of the Binder ratio
    in shown as a function of temperature $T$. A linear extrapolation to zero temperature
    is very well compatible with universality, i.e., with $g_c=1$.
    \label{binder_limit}}
\end{figure}

Another universal quantity is the value of the Binder ratio at the 
critical point $g_c$ which for the Gaussian model is $g_c=1$. For 
the discrete model the situation is slightly more involved because of
the cascade of fixed points in the RG flow. In the left panel of 
Fig.~\ref{binder_limit} we show the Binder ratio as a function of 
the system size (or number of generations $G$). It clearly shows the 
different scaling regimes. At small system sizes the discrete fixed 
point dominates the scaling behavior and increasing the system size 
initially leads to an increase of the value of the Binder ratio until,
upon increasing the size even further, we enter the scaling regime of 
the continuous fixed point where the Binder ratio starts to decrease 
when the system is made larger. The maximal value that the Binder ratio
reaches upon entering in the domain of the continuous fixed-point scaling
depends on the temperature as can be seen on the right panel of 
Fig.~\ref{binder_limit} and close to zero temperature it scales linearly
with a zero-temperature limit that is perfectly compatible with
$g_c=1$, i.e., that the value of the Binder ratio at the critical point
is also the same in both models. The fact that $g_c=1$ and that $\eta=0$
also means that there is one pair of physically relevant ground states 
in both models.

\section{Discussion}
Using the MKRG approach, we have studied 2D spin glasses, in
particular the model with discrete couplings and its two different
spin-glass phases at zero temperature. In the thermodynamic limit, one
can observe each of these phases depending on how the zero-temperature
limit is taken: If one takes {\it first} the zero-temperature limit
(or simply sets the temperature to zero right from the beginning as it
is often done in numerical ground state studies) and {\it then} sends
the size to infinity, the discrete fixed point is obtained. If
however, one {\it first} sends the size $L$ to infinity and {\it then}
sends the temperature to zero, then entropy fluctuations (and
temperature chaos) play a role and one is in fact observing a fixed
point identical to the one reached with continuous couplings, strongly
indicating that there is universality in 2D spin glasses. It is
important to note that the physically relevant fixed point is the one
which determines how quantities such as the spin-glass correlation
length or the spin-glass susceptibility diverge as the temperature
approaches the critical temperature $T_c=0$ and this is in all cases
the continuous-coupling fixed point. It is also important to note that
physically the domain-wall free-energy exponent is $\theta=1/\nu=-0.278$
also in the discrete coupling model if the thermodynamic limit is
taken such that the system size is always kept larger than $l_c(T)$
which notably is not the case if one simply sets the temperature to
zero in a finite-size scaling study (this leads to the well-known
$\theta^\prime=0$.) At finite size, the observed properties depend on
the length scale under probe and the crossovers between the different
regimes are associated with the chaotic length scale $\ell_c$ and the
paramagnetic one $L_{\rm eq}$.

The study of the continuous unstable fixed point is made difficult in
simulations because of the cascade in the MKRG flow. However, we have
shown that when using proper RG invariant quantities, one can measure
the exponents associated with the fixed points precisely. Let us
finally point out that this situation, where the entropy changes
drastically the properties at zero temperature, is not entirely
new. In fact, this is precisely the effect that changes the location
of the spin-glass transition in diluted 3D spin
glasses\cite{joerg:08b} from the naive percolation point to a
nontrivial one: this shows how much entropic effects, and temperature
chaos, matter in any renormalization or scaling scheme of
finite-dimensional spin glasses \cite{krzakala:00b}. In fact, such
entropic effects also matter in mean-field spin glasses and
optimization problems \cite{krzakala:07}. One has therefore always to
be cautious in using ground-state properties if the entropy is not
taken into account, because the presence of an additional unphysical
fixed point might lead to wrong results.

During the (long) preparation of this manuscript, we became aware of
the work of \cite{huse}, who first stressed the role of temperature
chaos. In fact the exponent $5.5$ they find in $2D$ is precisely
$\nu_{\rm eff}$ in Table 2. The recent simulations of
\cite{toldinPelissetto:10}, as well as the older ones in
\cite{joerg:06a}, display strong indications supporting our analysis
in 2D spin glasses on square lattices.

\section*{References}

\bibliographystyle{unsrt} \bibliography{../../refs}

\end{document}